**Geometry-Dependent Adhesion in Transparent, Monodomain Liquid Crystal Elastomers**

*Aidan Street*, Devesh Mistry, Johan Mattsson, and Helen F. Gleeson**

A. Street, D. Mistry, J. Mattsson, H.F. Gleeson
School of Physics and Astronomy, University of Leeds, Leeds, LS2 9JT, United Kingdom
E-mail: A. Street: py18ajs@leeds.ac.uk, H.F. Gleeson: H.F.Gleeson@leeds.ac.uk

ORCiD: A. Street: 0000-0002-7102-7644, D. Mistry: 0000-0003-0012-6781, J. Mattsson: 0000-0001-7057-2192, H.F. Gleeson: 0000-0002-7494-2100

Funding statement. AS is grateful for a studentship funded by Pilkington Technology Ltd. and the University of Leeds. HFG thanks the Engineering and Physical Sciences Research Council for a fellowship through EP/V054724/1. DM thanks the Leverhulme Trust for an Early Career Fellowship.

Data Availability Statement. The data collected during this study is freely able to be downloaded under the DOI https://doi.org/10.5518/1601.

# Geometry-Dependent Adhesion in Transparent, Monodomain Liquid Crystal Elastomers

Abstract: Elastomeric pressure-sensitive adhesives (PSAs) form adhesive bonds under light pressure. Liquid crystal elastomers (LCEs) are exciting PSA candidates as they can impart both anisotropy and temperature-dependence to adhesion, but the full potential of their anisotropic adhesion is unexplored. Here, identical side-chain LCEs, produced as transparent isotropic or nematic films are investigated; the latter aligned in homeotropic or planar geometries. Their room-temperature adhesion, determined through a 90˚ peel test, is consistent with theoretical predictions and strongest in a planar geometry (peeled parallel to the director) with adhesive force per unit length of 0.67N mm$^{-1}$. In contrast, adhesion of the planar perpendicular, isotropic and homeotropic films is 62.5%, 38.5% and 23.0% lower, respectively. The surface contribution to adhesion is identical for all films, confirming that the variation in adhesion is determined solely by the bulk LCE alignment controlled during film preparation. A temperature-dependent adhesion factor is determined from 0 ˚C to 80 ˚C using dynamic mechanical analysis, and found to be in excellent agreement with the peel data at room temperature. Molecular relaxations active above the glass transition temperature are dominant in determining LCE adhesion. The results show that side-chain LCEs can function as transparent, tunable, broad-temperature smart PSAs.

1. Introduction

Elastomers with adhesive properties (such as Gorilla Tape $^{TM}$) are attractive for applications, since they can maintain adhesion while also allowing deformations. Elastomers characterized by a viscoelastic room temperature response, thus demonstrating adhesion upon application of light pressure, constitute a highly desirable class of pressure-sensitive adhesives (PSAs), [1] and since elastomeric adhesives do not flow, they are particularly useful in joints or vibrating components that rely on flexural changes or undergo significant temperature changes. [1] For certain applications, including pressure-sensitive tapes and no-sealant gaskets, more advanced adhesives composed of chemically identical components that debond at different loads, are desired. [2] An emerging class of PSAs that can offer both directional and temperature-dependent adhesion is based on liquid crystal elastomers (LCEs), for which there has been particular focus on thermal actuation. In the most recent reports, low-temperature adhesion in LCEs is sufficiently strong to easily out-compete some incumbent PSAs, with facile release at high temperatures. [3,4,5,6] However, there has been only one experimental study that considers anisotropic adhesion in LCEs, offering strong adhesion in one direction with low adhesion in another, enabling easy removal. [4] In this paper, we explore the adhesion of chemically identical, monodomain films of an LCE formed in all four possible geometries, allowing us to determine the key factors in their performance as smart pressure-sensitive adhesives. We also consider the temperature-dependence of the adhesion strength in all cases, suggesting application areas most relevant to this class of smart adhesives.

LCEs combine the long-range orientational order of a liquid crystalline material (along a director, $\hat{n}$) with the viscoelastic properties of a rubbery elastomeric material. [7] The mesogenic units of the LCE can be either part of the polymeric backbone (main-chain LCE) or linked to it *via* a spacer unit (side-chain LCE). Monodomain LCEs, in which the director is aligned macroscopically, are inherently anisotropic (see Figure 1(a) for the different possibilities) and it is this property that offers the directionality of adhesive strength. However, it is the opportunity to tune the liquid crystal phase range in LCEs that has so far been taken advantage of in temperature-sensitive PSAs. Here, we consider both the directionality and temperature-dependence of a model family of LCEs that can be prepared in various forms, allowing us to test theoretical predictions and enhance the current understanding of LCEs as PSAs through the simultaneous consideration of relevant material parameters.

The debonding process that occurs in LCEs was first considered theoretically by Corbett and Adams who simulated the adhesive properties of an LCE under a probe-tack test, following a methodology introduced by Gay and Leibler for isotropic soft polymer adhesives. [8] The Corbett and Adams work considered the LCE to comprise of blocks that can experience stretch, shear and slippage. [8] They predicted that ordering of the LCE is important; when the alignment of the director is parallel to the detachment direction, more than double the debonding energy is expected compared with the unaligned (isotropic) LCE. [8] The isotropic LCE itself is predicted to possess a debonding energy that is more than twice as large as when the nematic director is oriented perpendicular to the detachment direction. [8]

Most of the reports that examine LCEs as PSAs have focused on the difference between adhesion in the nematic and high-temperature isotropic states of polydomain main-chain LCEs. Ohzono *et al*. found that adhesion in the nematic phase was double that in the high-temperature isotropic phase for the first LCEs they studied, a difference attributed to the increased energy dissipation of the nematic phase during the detachment process. [3] Nematic phase adhesion as high as ~2000 Nm$^{-1}$ has since been achieved by inclusion of polyrotaxanes in the LCE; [6] this is more than 5x larger than the 3M 309 tape, a polypropylene tape typically used as for packaging, and the LCE adhesion decreases to ~50 Nm$^{-1}$ at 100 ˚C, close to the isotropic phase, allowing easy thermally activated debonding. The work of Terentjev and co-workers on LCEs as novel temperature-controlled command adhesives, [3,6,9,10] has shown a correlation between the mechanical loss factor $\tan \delta = E''/E'$ and the adhesive energy of LCEs, confirming that the dissipative properties of the nematic materials are extremely important. [3,9] The polydomain geometry takes advantage of the nematic nature of the LCE in optimizing the PSA, but confers no anisotropy on the film. However, Pranda *et al*. demonstrated the potential of anisotropic adhesion, showing that an aligned, main-chain LCE-based adhesive can exhibit up to 9 times higher adhesive force perpendicular to the director than parallel to it when subjected to 180˚ peel tests at a speed of 0.5 mms$^{-1}$. [4] Pranda *et al* also correlated a stronger adhesion with an increase in $\tan \delta = E''/E'$ and showed enhanced adhesion close to the glass transition. [4] The work of Pranda *et al* represents the first consideration of LCEs as anisotropic PSAs, but examines only two of the four possible film geometries that could potentially be achieved with a specific elastomer.

In our work, we investigate a set of chemically identical side-chain elastomers, manufactured as monodomain films that are either isotropic or nematic, the latter with homeotropic or planar alignment, allowing us to investigate all four possible geometries, as shown in Figure 1 (a). We note that there are two cases where the nematic director is perpendicular to the detachment direction, homeotropic and planar ⊥ (Figures 1 (a) and (c) respectively), and we consider these separately. This nematic LCE has been widely studied for its auxetic response which occurs at strains typically ≳ 0.4, though this strain is much higher than experienced in the debonding experiments reported here. The materials were chosen because it is straightforward to produce highly transparent (see Figure 1(b)), monodomain films with high macroscopic order (nematic order parameter ~0.6) and a glass transition temperature below room temperature ($T_g \sim 15°C$).[11,12] Importantly, a chemically identical isotropic film can also be formed with a similar $T_g$ allowing us to deconvolute the influence of the phase of the film (whether nematic or isotropic) from the contribution of the bulk moduli to adhesion of the samples, the latter varying most dramatically for any polymer near the glass transition temperature. All the films we consider have equivalent surface energies ensuring that the surface contribution to tack can be disregarded in their comparison.

2. Results and Discussion

We determine the room-temperature adhesion of the elastomer films in geometries shown schematically in Figure 1 through a 90° peel test where no cohesive failure was observed.[13] Since the surface bonding is important in determining adhesion, we first confirm that the surface energy, assessed by determining the contact angle formed by a freestanding drop of water, is equivalent for all the films.[14] We then consider the temperature-dependence of the adhesion through measurements of the mechanical material response, as the relevant mechanical moduli can be directly used to predict the adhesive strength across a temperature range of interest.[15,16]

(a) (b)

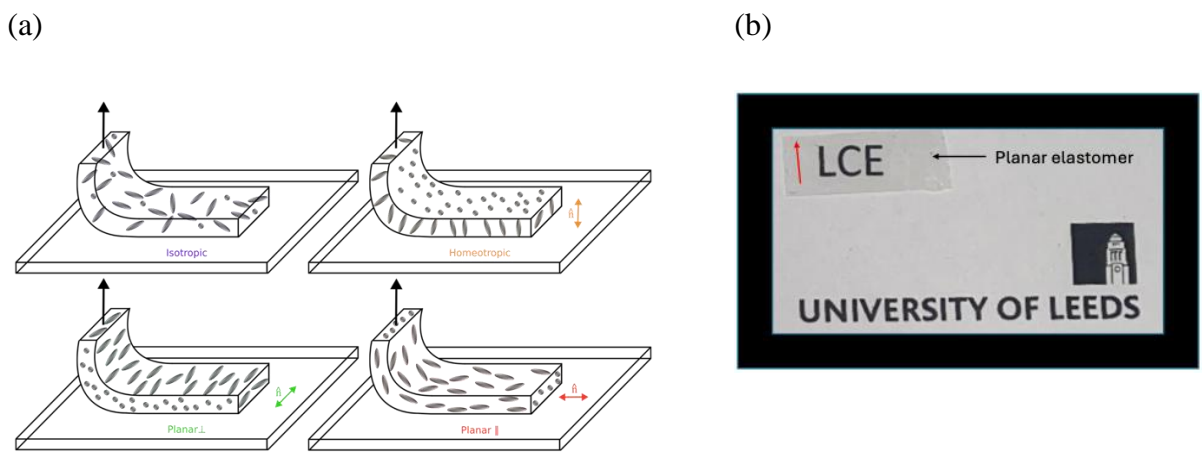

Figure 1. (a) Idealized schematic of the relative peeling directions of the various LCE films with respect to the glass substrate (top-left) isotropic, (top-right) homeotropic, (bottom) planar with the director perpendicular and parallel to the peel direction from left to right respectively. Note that in both the homeotropic and planar ⊥ cases the director is perpendicular to the peel

direction (black arrow). (b) Photograph of a sample of the planar LCE film, demonstrating the high transparency which has been quantified in [11]. The director is indicated by the red arrow, along the short sample edge.

The behavior of an adhesive in a peel geometry can be described theoretically [15] and the debonding force ($F$) per unit width ($w$) is simply related to the adhesive failure energy per unit area ($\Theta$) by the cosine of the peel angle $\gamma$ through Equation 1

$$\boldsymbol{\Theta = (1 - \cos(\gamma))\frac{F}{w}}. \qquad \textbf{(Equation 1)}$$

This assumes that the peeling arm does not undergo any tensile deformation. As a constant 90° peel angle is maintained in our experiments, $\cos(\gamma)$ is zero, and a factor of 2 is expected between our approach and that of Pranda *et al* who employ a 180° peel. Importantly, $\Theta$ can be described as the sum of an interfacial bond energy ($\boldsymbol{\Theta_0}$) and a bulk term $\Delta\Theta$, i.e. $\boldsymbol{\Theta = \Theta_0 + \Delta\Theta}$. The bulk term is given by a function $\phi(v, T, \epsilon)$, where $v$ is the peel velocity, $T$ is the temperature, and $\epsilon$ is the strain. [15] Consequently, we can separately analyze the surface and bulk contributions to the adhesion.

2.1. Surface Energy Considerations

The interfacial bond energy is governed by the surface energy of the materials in contact. [14] To probe the relative magnitude of surface energy of the LCE films as a function of strain, thereby considering its contribution to adhesion for all the experiments undertaken, we performed contact angle measurements at room temperature (above $\boldsymbol{T_g}$) to assess the wetting of a 3 µL drop of deionised water. The contact angle was measured up to strains of 1.0, which is far beyond the maximum strain experienced by the LCE films during the adhesion tests which is less than ~0.08.

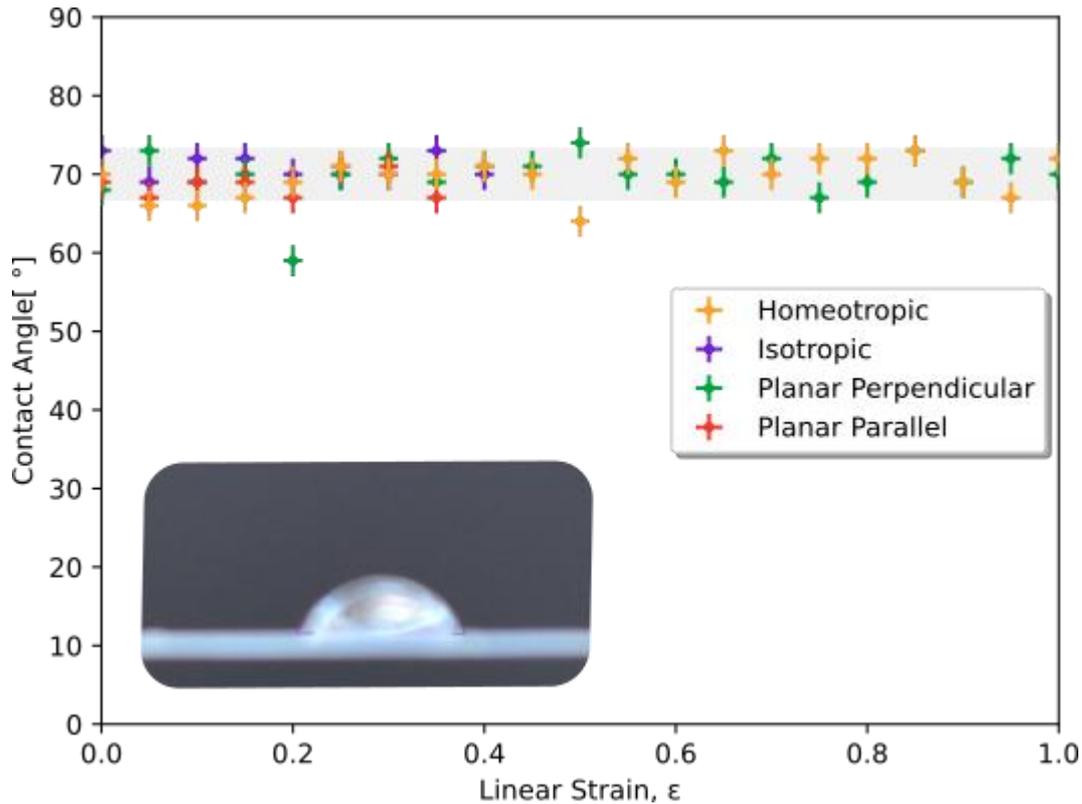

Figure 2. Contact angle of water as a function of strain for the LCE systems: homeotropic (orange), isotropic (purple), planar ⊥ (green), and planar ∥ (red). The wetting angle is unchanged for all samples and linear strains. Inset, the contact angle of an individual 3 μL drop on the isotropic LCE.

As shown in Figure 2, the contact angle is constant within experimental uncertainty for all four film geometries and is independent of strain, in agreement with expectations. Previous work on elastomers has shown that in the glassy state ($T < T_g$), the surface energy is highly strain dependent, manifesting in large-scale changes in contact angle with applied strain. [18] However, in the non-glassy rubbery state ($T > T_g$), as in our experiment, the molecular motion on length-scales significantly less than the average inter-crosslink distance, is fluid-like and no strain-dependence is typically observed. [18] The fact that no differences in contact angles are observed for our four LCE geometries suggests that the contribution of the interfacial bond energy to Θ can be disregarded with respect to any observed differences in adhesion. This means that changes in Θ are due to changes in the LCE bulk properties.

2.2. Adhesion as Measured by Perpendicular Peel Tests
In our work, the peel test involves peeling the adhesive away from a rigid glass substrate, where the peel force is applied parallel to the substrate normal. [4] To ensure that a relevant, commonly encountered, adhesion situation was probed, no heat treatment was used to assist in the bonding and the peel tests were performed immediately after the films were adhered to the substrates. Figure 3 shows the average force per unit width required to debond each of the four elastomer films from the glass substrate. For nematic films, clear differences are observed in the peel force depending on the alignment of the director relative to the peel direction. The homeotropically aligned LCE requires the least force to debond, while the nematic planar ∥ sample requires more than four times the homeotropic-case force. Interestingly, the forces for the homeotropic and planar ⊥ films differ by roughly a factor of three even though the peel is perpendicular to the director in both cases and they have the same elastic modulus. [12] Specifically, an average debonding force per unit width of 0.067 Nmm$^{-1}$ is required when the director is aligned with the peel direction (planar ∥) compared with 0.043 Nmm$^{-1}$ for the planar ⊥ geometry and 0.015 Nmm$^{-1}$ for the homeotropic geometry, respectively. The force per unit width required to debond the isotropic film lies between the values for the homeotropic and planar ⊥ (0.026 Nmm$^{-1}$) geometries.

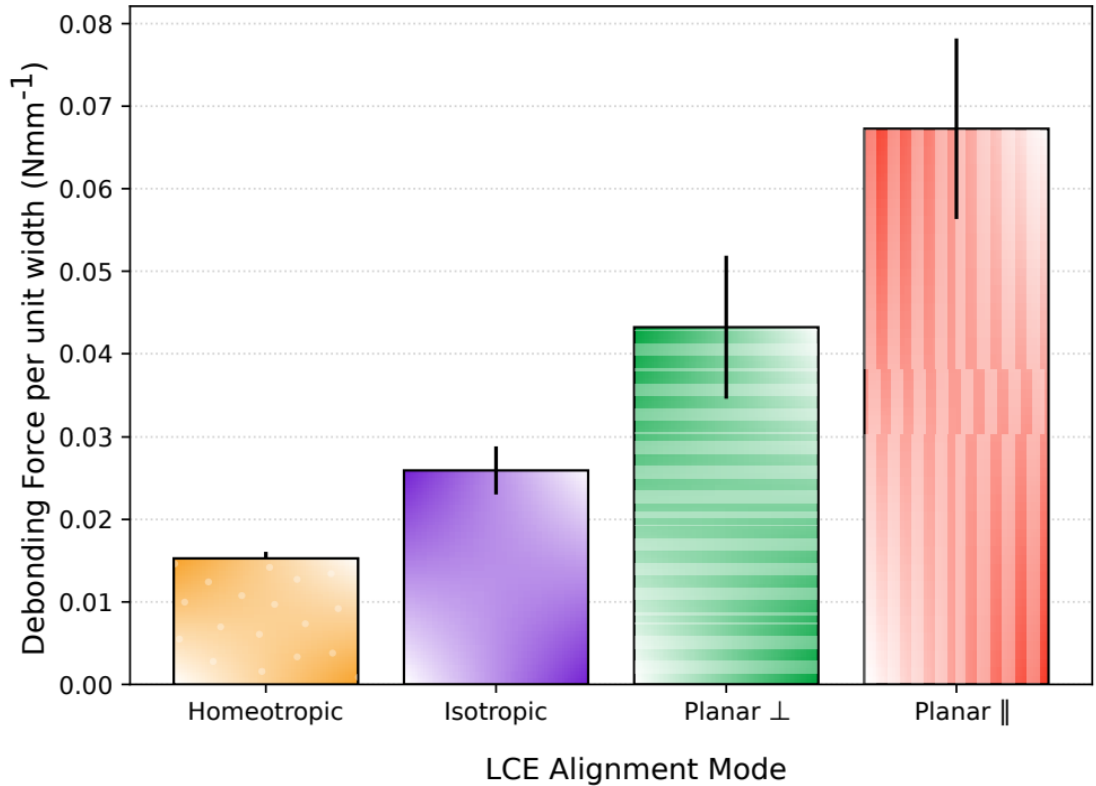

Figure 3. Force per unit width required to debond the LCE from the glass substrate at a peel angle of 90° and at room temperature (20.5 °C) for the isotropic (purple), homeotropic (orange), planar ⊥ (green) and planar ∥ (red) monodomain films. The statistical variance is indicated by the error bars.

| Geometry | This work (experiment) | Corbett and Adams (simulation) [8] | Pranda et al. (experiment) [4] | | |
|---|---|---|---|---|---|
| Planar ∥ | 4.4 | 4 | 0.1 | 0.3 | 0.71 |
| Planar ⊥ | 2.9 | - | 1 | 1 | 1 |
| Homeotropic | 1 | 1 | - | - | - |
| Isotropic | 1.73 | 2 | - | - | - |

Table 1. Relative adhesive strength of LCE geometries, normalized to the homeotropic case, also considered theoretically by Corbett and Adams. Note that Pranda *et al* did not consider either homeotropic or isotropic geometries so their data are normalized to the planar ⊥ case

allowing a simple comparison of materials they considered with different molar content of crosslinker (0.5, 0.7 and 0.9 left to right).

Our results can be directly compared with experimental results from Pranda et al and simulations by Corbett and Adams, see Table 1. The nematic film in the simulations was assumed to have an order parameter of 0.7, rather large for nematic LCEs, but comparable to the value of ~0.6 for both our films and those reported by Pranda *et al*. [4] The simulation found a ratio of 2.07:1 between the planar ∥ and the isotropic cases, and a ratio of 0.41:1 between the homeotropic and isotropic cases. These numerical results are in surprisingly good agreement with our data for which the planar ∥ alignment shows the greatest adhesion with a ratio of ~2.6:1 to the isotropic case, and a ratio of ~0.6:1 between the homeotropic and isotropic case. Interestingly, for the peel test measurements performed by Pranda *et al*. (for a main-chain LCE in planar ⊥ and planar ∥ geometries), [4] the debonding force was higher in the planar ⊥ configuration, a result that is opposite to our observations. The difference between which geometry has the maximum adhesion is intriguing and cannot be explained by the fact that our peel tests were at 90°, while those of Pranda *et al* were 180°, nor can it be attributed to difference in Young's moduli which are ~10 MPa for both our films and the main-chain systems considered by Pranda *et al*. [4,19] Perhaps the most important difference is that in our work, the maximum strain did not exceed ~0.08, so the LCE deformation was always in the initial elastic regime, while Pranda *et al* mention the influence of the soft elastic regime (which occurs at relatively higher strains) on the adhesive properties of their LCEs.

2.3. Adhesion Factor from Dynamic Mechanical Analysis
To fully investigate the link between the observed adhesive behavior and the bulk properties of the LCE, [1] we now focus on the relationship between the adhesive properties and the mechanical shear moduli across a wide temperature range. Small-amplitude oscillatory shear experiments were performed using dynamic mechanical analysis (DMA), as outlined in detail in the Experimental Section. For each of the four geometries, two equivalent LCE samples were each placed between the fixed outer plate and the middle moving plate of the DMA shear tool, as shown in Figure 4(a). By oscillating the middle plate of the tool, a sinusoidal stress is applied and the resulting sinusoidal deformation is measured, and thus the frequency-dependent complex shear modulus, $\boldsymbol{G^*(\omega) = G'(\omega) + iG''(\omega)}$ can be determined, where the shear storage modulus $\boldsymbol{G'(\omega)}$ is the in-phase component and the shear loss modulus $\boldsymbol{G''(\omega)}$ is the out-of-phase component. The ratio between the moduli, the loss tangent, $\boldsymbol{\tan\delta(\omega) = G''(\omega)/G'(\omega)}$, provides a measure of the dissipation at the probed angular frequency ω. We determined $\boldsymbol{G^*(\omega)}$ and thus $\boldsymbol{\tan\delta(\omega)}$ for a probe frequency of ω = 1 rads$^{-1}$ across a temperature range from 0 °C to 80 °C, see Figure 4. To note, much of the work in literature uses the tensile modulus $\boldsymbol{E'}$ for LCEs; for isotropic systems, the value for $\boldsymbol{G'}$ differs by a factor of $\boldsymbol{2(1+\nu)}$ where $\boldsymbol{\nu}$ is the Poisson's ratio. [22]

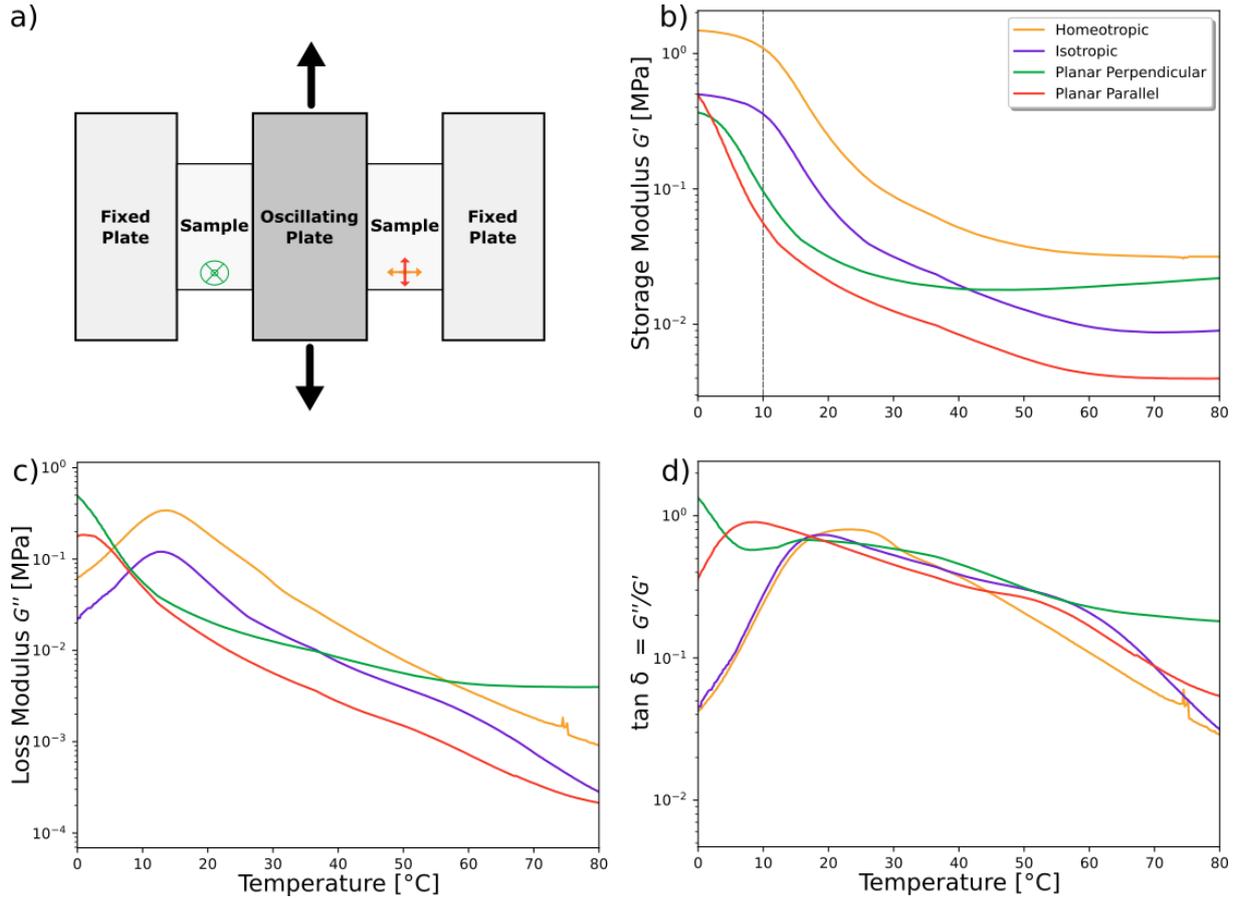

Figure 4. Dynamic mechanical analysis results for the LCE films. (a) a schematic representation of the films sandwiched between steel plates, illustrating the alignment relative to the shear direction, as shown for homeotropic (orange), planar ⊥ (green), and planar ∥ (red) (the isotropic case is unaligned). (b) the storage modulus $\boldsymbol{G'}$, (c) the loss modulus $\boldsymbol{G''}$ and (d) the loss tangent, i.e. the ratio between them $\boldsymbol{\tan\delta = G''/G'}$. The dashed line in (b) indicates the glass transition temperature, $T_g$.

Figures 4 (b-d) show the shear properties of the films from 0°C to 80°C. In general, the storage moduli ($\boldsymbol{G'}$) show two separate features corresponding to two different relaxation modes in the LCE, which manifest as peaks in the loss moduli ($\boldsymbol{G''}$) at low and higher temperatures; the features are least obvious in the planar ⊥ geometry. The low-temperature mode occurring at ~10 °C is related to the structural α-relaxation corresponding to the glass transition. The two relaxation modes are also manifested in the loss tangents, where the values largely overlap at a value of tan δ ≈ 0.6 and decrease similarly as the temperature increases. At room temperature (20.5 °C), the value taken by tan δ is within 18% for all the films. The higher temperature relaxation demonstrates the presence of an additional, slower relaxation mode in these LCEs. [22] There are two peaks present in the loss tangent (tan δ) and the contribution of the second (high temperature) peak is weaker. Such behavior is commonplace among other LCEs. [3] Interestingly, this additional high-temperature relaxation is more significant than in a similar formulation of this LCE with twice the mol % of crosslinker, [21] for which we suggest that the highly crosslinked nature prohibited some relaxation modes in the bulk. We find that the planar parallel LCE showed the lowest $\boldsymbol{G'}$, i.e. the softest response, and the homeotropic LCE the largest $\boldsymbol{G'}$, i.e. the stiffest response.

Interestingly, these results contrast literature results for the tensile moduli of many LCEs, which for geometries where the strain is perpendicular to the pre-strain director there is a much softer response due to director rotation associated with their uniaxial (semi-soft elastic) response.

It is clear that the tan δ values alone do not account for the measured differences in adhesion of the LCE alignment systems from the peel tests presented earlier. [22] An adhesion factor, $\mathcal{A}$, defined as the ratio of the loss tangent (tan δ; damping factor) to the storage modulus $G'$, provides a metric by which to compare the adhesion of viscoelastic materials (Equation 2). [3,8]

$$\mathcal{A}(\omega) = \frac{tan\delta(\omega)}{G'(\omega)} = \frac{G''(\omega)}{(G'(\omega))^2} \qquad \textbf{(Equation 2)}$$

A high value of $\mathcal{A}$ indicates a material that can be readily deformed whilst simultaneously displaying a high energy dissipation, and this has been used successfully to evaluate LCE systems. [3] $\mathcal{A}$ can be determined for a fixed probe frequency (typically ~1 rads$^{-1}$) as an approximation of the conditions of a typical adhesion test. [3,8] The values of $\mathcal{A}$ for the four LCE geometries, calculated using the data shown in Figure 4, are presented in Figure 5. As expected, due to the two peaks observed in tan δ, two peaks are also clearly visible in $\mathcal{A}$, except for the planar ⊥ alignment, for which the asymmetric peak shape suggests the presence of a second weaker overlapping peak. [8] At room temperature, the values of $\mathcal{A}$ are in ascending order: homeotropic: 3.4 MPa$^{-1}$; isotropic: 10.1 MPa$^{-1}$; planar ⊥: 21.5 MPa$^{-1}$; and planar ∥: 31.3 MPa$^{-1}$. The corresponding values of the adhesive force per unit width from the room temperature peel test experiments are shown in the filled circles and clearly demonstrate an excellent correlation between the two approaches.

We note that determining $\mathcal{A}$ in this way allows for facile predictions of the adhesion properties of these materials as PSAs over a wide temperature range. Interestingly, for this material, the adhesion factor continues to increase with temperature for the planar parallel and isotropic cases above the α-relaxation regime due to the presence of the second high-temperature relaxation. This has an effect on the relative values of $\mathcal{A}$ for the different geometries, for instance for T > 45 °C, the value for the isotropic phase becomes larger than that for the planar ⊥ geometry, implying that a larger adhesive force is required to peel the corresponding LCE off a substrate. Such an observation is unusual for LCEs where it is most often suggested that the isotropic state will have a lower adhesion than the nematic state; clearly that will only be the case definitively if the temperature-dependence of the adhesion is dominated by the α-relaxations and the isotropic state is accessed by increasing temperature.

The highest anisotropy in the adhesion factor for the two planar geometries is found at 52 °C, where the $\mathcal{A}$ value for the planar parallel geometry is twelve times higher than the perpendicular case (48.2 MPa$^{-1}$ compared to 3.9 MPa$^{-1}$). This ratio is comparable to the corresponding ratio (~9) found for main-chain LCEs by Pranda *et al*, but in our case the higher adhesion is found for the planar parallel case. This confirms that side-chain LCEs can offer similarly high anisotropic adhesion as was observed for the main chain system in Pranda *et al*'s study. 𝒜 has also been determined by Ohzono et al. for a polydomain main-chain LCE. [3] The LCE in that work underwent a nematic-isotropic transition just above room temperature and they found that 𝒜 was a factor of around 2.5 lower high in the isotropic phase (60 °C) than the maximum in the low-temperature nematic phase.

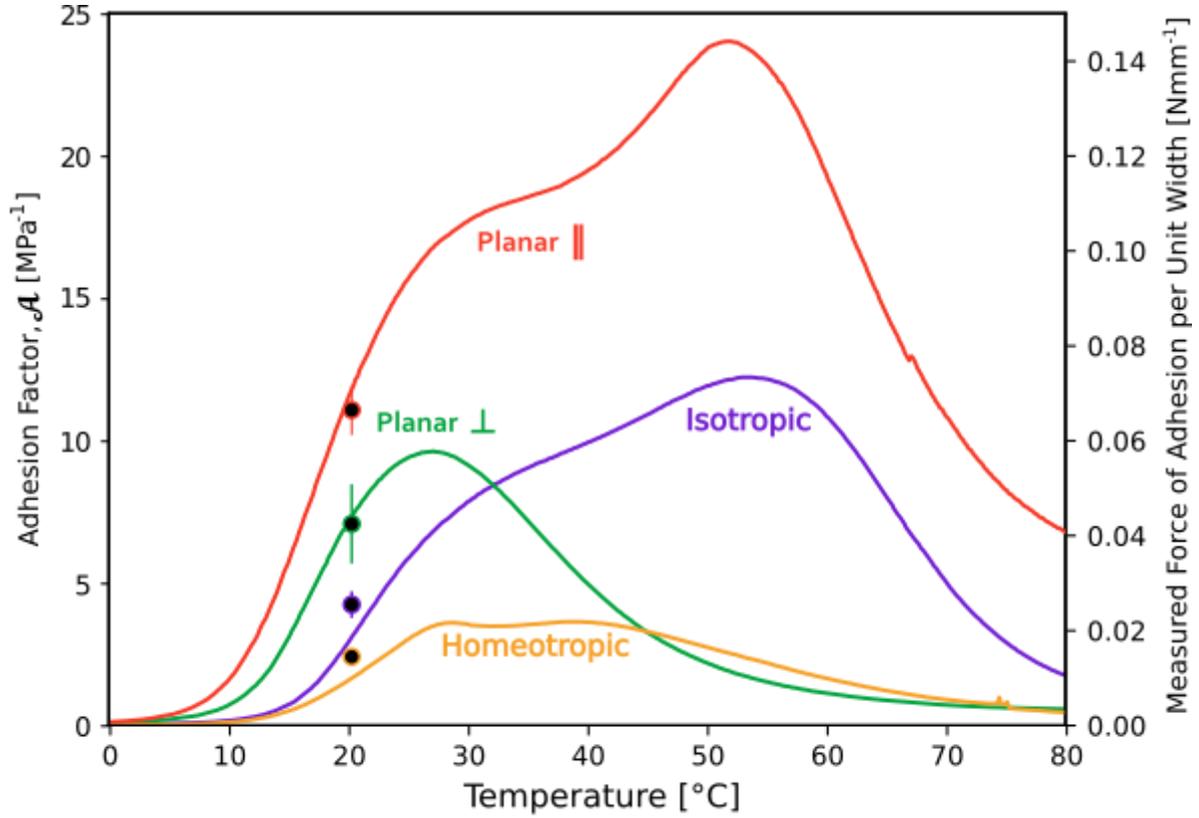

Figure 5. The adhesion factor, $\mathcal{A} = \tan\delta/G'$, determined at ω = 1 rads$^{-1}$, as a function of temperature for the various samples: homeotropic (orange), isotropic (purple), planar ⊥ (green), and planar ∥ (red). The black circles superimposed at 20.5 °C show the adhesion force per unit width for each of the samples measured at using the 90° peel test.

2.4. Practical Adhesive Uses for LCEs

LCEs are increasingly competitive with commercial PSAs and while optimising the adhesive properties was not a consideration in the material selection for this work, it is nonetheless interesting to consider potential applications of the materials. In addition to the adhesion factor, the Dahlquist criterion, which considers the potential performance of a PSA from shear moduli is often used to define an 'ideal' PSA. [24] The Dahlquist criterion, $G_C$, represents a modulus value for which the PSA is able to spontaneously adjust to a surface of a given roughness; the adjustment of the PSA to the surface corrugation increases the contact area and thus the work required to remove the adhesive from the surface. [24] $G_C$ can be defined as Equation 3:

$$G_C = W\sqrt{\frac{R}{h^3}}, \qquad \text{(Equation 3)}$$

where, $W$ represents the work of adhesion between the PSA and the surface, and $R$ and $h$ respectively are the average radius and height of a bumpy feature on the surface. Typical values of $W$ = 50 mJm$^{-2}$, $R$ = 5 μm, and $h$ = 1 μm, suggest that $G_C$ ~ 0.1 MPa and this, together with the requirement of a low $T_g$ (~ 0 ˚C), has been used in defining an ideal PSA [10]. Despite not being selected for their adhesive potential, the LCEs in this work begin to approach these ideal conditions, with values of $G'$ ~ 0.08 MPa at a temperature of 18.9 ˚C and $T_g$ ~ 10 ˚C in the planar ∥ case.

An alternative measure employs the so-called Chang plot (see Fig. 6) which extends the Dahlquist criterion by considering the storage and loss moduli of a material. [25] The Chang plot is divided into four quadrants that indicate typical real-world uses of PSAs as shown in Figure 6 where also the Dahlquist criterion is marked ($G' = G_C$). In the plot, different adhesives fall into different quadrants: classic adhesive (upper-left); high shear (upper-right); removable (lower-left); and cold temperature (lower-right) PSAs. [25] The Chang graph is typically plotted at room temperature using data measured at different frequencies, although we have here used it to visualize the working temperatures for which our films might be candidates to address specific tasks. Below $T_g \approx 10$ ˚C all our materials have storage moduli above or close to $G_C$, indicating that they are not ideal PSAs. Above $T_g$ the LCEs are initially all closely grouped into the high shear PSA section of the plot, and as the temperature increases further the sample properties spread within the lower left quadrant. This suggests possible applications of the LCEs in this work as the homeotropic LCE in high-shear scenarios, typically in outdoor environments. [25] Those LCEs that belong to the lower left 'removable' area have a high contact efficiency but low dissipation, making them well-suited to applications such as masking. At 37 ˚C the planar parallel LCE is the furthest into the 'removable' area, this makes it particularly fitting for possible use as a medical tape. [25] It is worth noting that all of the films considered here are highly transparent, a property not shared by polydomain LCEs and a feature that extends their potential use cases.[11]

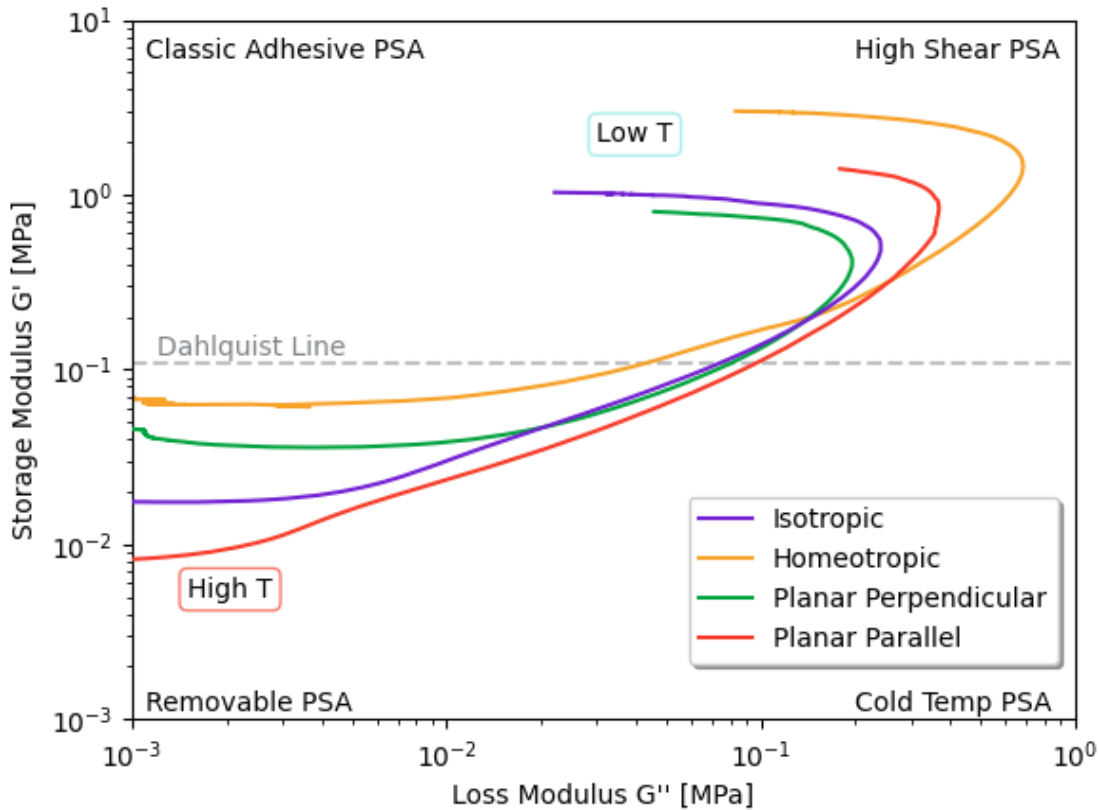

Figure 6. The Chang plot, $G'(\omega)$ as a function of $G''(\omega)$. The coloured lines show the temperature-dependent moduli of our samples which shift through three of the quadrants as the temperature is varied. The positions of the isotropic (purple), homeotropic (orange), planar ⊥ (green), and planar ∥ LCE samples are shown for the temperature range 0 ˚C to 80 ˚C. At low temperatures (below $T_g$) the LCEs are above the Dahlquist line ($G' = 0.11$ MPa) and therefore are not recommended for general adhesive applications.

3. Conclusions

This paper considers both the anisotropic and temperature-dependent adhesive properties of chemically identical side-chain elastomeric films that are nematic or isotropic, with all possible geometries examined in the former case. In all cases, the LCE detached from the glass substrate through adhesive rather than cohesive failure. Our results indicate that the bulk mesogenic alignment of LCEs greatly affects the adhesive strength of the elastomers to glass. At room temperature, a minimum adhesive energy (0.015 Nmm$^{-1}$) was observed for the homeotropically aligned LCE, where the mesogens are aligned 'end-on' to the adhesive interface, while the maximum adhesive energy (0.67 Nmm$^{-1}$) was recorded for the LCE with planar alignment parallel to the detachment interface. The isotropic film had intermediate adhesive energy (0.026 Nmm$^{-1}$). These findings are in excellent qualitative and quantitative agreement with the predictions of Corbett and Adams [8].

Strong anisotropy was observed in the adhesive strength for the planar parallel and planar perpendicular orientations. While significant adhesion anisotropy for planar LCEs has been reported previously, the geometry with the maximum adhesion differs; it was planar parallel in our case and planar perpendicular for Pranda et al. The cross-over in anisotropy cannot be attributed to differences in the order parameter or moduli of the two systems, which are very similar. However, there are differences in both the LCE type (side-chain in this work and main-chain in Pranda *et al*) and in the maximum strain applied for debonding. In the work of Pranda *et al*, it was suggested that the strain was sufficient to take the planar perpendicular LCE into the semi-soft regime (this is not possible for the planar parallel case). This was not the case for our LCE system; it does not exhibit a semi-soft deformation and the maximum strain was 0.08, far less than is required for any significant director reorientation.

A temperature-dependent adhesive factor was determined for the films via dynamic mechanical analysis. The LCE material considered here shows two relaxations as a function of temperature, the *α*-relaxation associated with the glass transition in addition to a higher-temperature relaxation. This gave an unusual variation of the adhesion factor which was not monotonic with temperature and the occurrence of the second relaxation peak meant that the relative magnitude of adhesion changed for the different geometries at higher temperatures, though the planar parallel was always the greatest. A maximum directional difference of a factor of 12 is predicted for this system between the two planar geometries at a temperature of 52 ˚C.

This study of chemically identical, optically clear LCEs with different film geometries has demonstrated that both the temperature-dependent adhesion and its anisotropy can be tailored for desired purposes solely through bulk mesogenic alignment. It is facile to pattern these films through a combination of surface alignment, externally applied fields and temperature, suggesting programmable adhesive strips.[27, 28] For this system, the potential use areas of the films as adhesives changes with temperature, as shown by superimposing the data onto a Chang plot. We note that there is scope to adjust the adhesive failure energy further, through chemical means, including changes in cross-link density, to shift the position of $T_g$ and alter the shear moduli. [16] In this way the adhesive strength of the LCE can be tuned and optimized further, depending on the desired application.

4. Experimental

LCE Preparation and Characterization

The formulation of liquid crystal elastomer used in this work has been reported in detail previously, but a brief description is provided here. [26] A conventional liquid crystal mold was produced using one glass substrate and one flexible polymer substrate, with 100 μm thick strips of Melinex (DuPont Teijin films, Japan) used as spacers (Figure 7). For planar aligned LCEs, the interior surfaces of the substrates were spin-coated with a 0.5 wt.% polyvinyl alcohol solution and uniaxially rubbed once dry. For homeotropic aligned LCEs, an indium-tin oxide (ITO) coated glass substrate (Xinyan Technology Ltd., People's Republic of China) and an ITO coated PET film (Sigma-Aldrich, United States of America) were spin-coated with a 0.5wt. % cetyltrimethy-lammonium bromide (MP Biomedicals, France) and remained unrubbed.

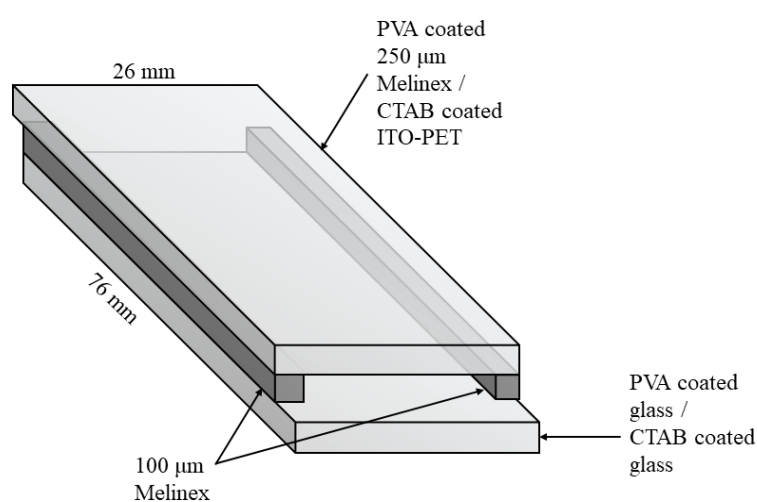

Figure 7. A schematic of the mold used for LCE production. The precursor mixture was filled into the (76 mm x 26 mm x 100 μm) mold *via* capillary action in the isotropic phase. For the homeotropic sample, a voltage of 40 $V_{rms}$ at 1 kHz was applied across the sample (perpendicular to the substrate) to augment the surface alignment.

A precursor mixture was produced consisting of; a monofunctional mesogenic side group 6-(4-cyano-biphenyl-4'-yloxy)hexyl acrylate (A6OCB), a bifunctional mesogenic cross-linker, 1,4,-bis-[4-(6-acryloyloxyhexyloxy)-benzoyloxy]-2-methylbenzene (RM82), an ultraviolet photoinitiator methyl benzoylformate (MBF), a non-mesogenic unit 2-ethylhexyl acrylate (EHA), and the non-reactive mesogenic material 4-cyano-4'-hexoxybiphenyl (6OCB). The structures of all of the materials used are shown in Figure 8. The non-reactive mesogen, 6OCB, is included in the precursor mixture in order to extend the nematic phase range prior to polymerization where a nematic LCE film is the final product; it is washed out of the final LCE. [11] A6OCB, 6OCB, and RM82 were supplied by Synthon Chemical GmbH (Germany), whilst MBF and EHA were obtained from Sigma Aldrich (United Kingdom).

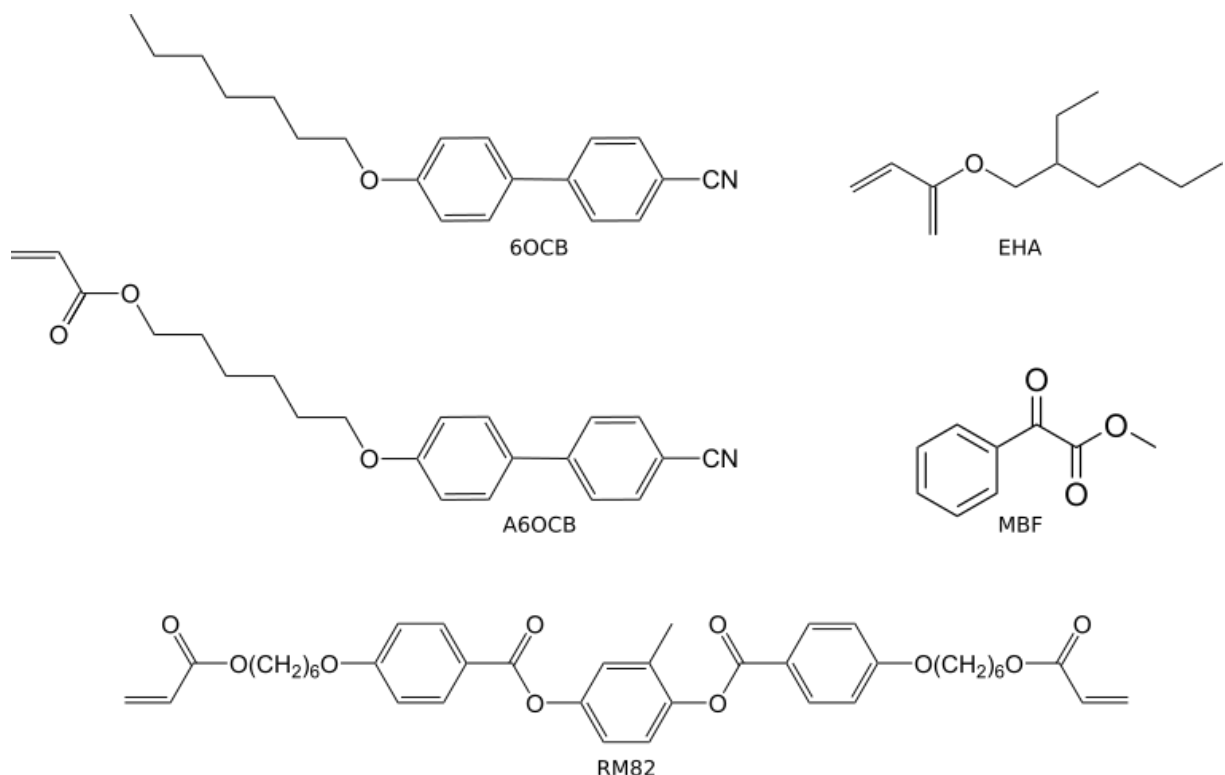

Figure 8. Structures of the LCE precursor components, the 6OCB is washed out of the final LCE film.

The mesogenic materials were heated to 120 °C and stirred for 5 min, the mixture was cooled to 40 °C before the non-mesogenic materials MBF and EHA were added via pipet.
The mixture was stirred for a further 5 min before being filled into the molds via capillary action. To form the nematic LCEs, the molds including the mixtures were then cooled to the nematic phase (room temperature) before a low intensity 2.5 mWcm$^{-2}$ ultraviolet light source was used to cure the acrylate polymer. For homeotropic LCEs, a voltage of
40 V$_{rms}$ at 1 kHz was applied across the mold to augment the surface alignment before and during the cure. Isotropic LCEs were formed by curing at an elevated temperature of 40 °C, in the isotropic phase of the precursor mixture. Post-cure, molds were disassembled by peeling off the Melinex substrate and the LCE films were placed in a 30% dichloromethane in methanol mixture overnight to remove the 6OCB. The LCEs were then left in ambient conditions for 2 h to dry. Table 1 shows the wt. % of the components before and after the washing step. The resultant aligned LCE films are optically clear and conserve volume under a deformation. [11,20] This method produces monodomain, optically clear LCE films of dimensions 7 cm x 2.5 cm and thickness ∼ 100 μm that are isotropic, or nematic with planar or homeotropic alignment; the optical properties of the films have been reported previously.[11, 12] The nematic nature of this LCE has been confirmed via x-ray scattering, and the order parameter has been measured using Raman spectroscopy to be $\langle P_{200} \rangle \sim 0.60$. [12] This family of nematic LCEs show no phase transition to the isotropic phase below 100˚C. [11,21]

| Component | Precursor Mixture [mol%] | LCE [mol%] |
| --- | --- | --- |
| 6OCB | 54.6 | - |
| A6OCB | 24.4 | 53.8 |
| EHA | 16.0 | 35.2 |
| RM82 | 3.5 | 7.7 |
| MBF | 1.5 | 3.3 |

Table 2. Pre-wash and post-wash formulation of the LCE. The chemicals are identified in Figure 8.

The thermal properties of this nematic LCE have been rigorously studied previously through differential scanning calorimetry. [11,12] The onset of the glass transition peak on cooling determined to be $T_g \approx 10$ ˚C with no discontinuous $T_{ni}$ found below T = 200 ˚C. [11]

Peel Tests
The adhesive properties of the LCES were determined using a 90˚ peel test as indicated in Figure 1. To assess the adhesive strength of the LCE to glass, an approximately 10 mm x 7 mm strip was pressed onto a glass slide using a force of 3.9 kN for a period of 5 min. The ensemble was suspended upside down and a variable mass was hung from the strip. The suspended mass was increased until a steady state peel was achieved; a peel angle of 90˚ was maintained. Each test was performed five times and the standard deviation used in error analysis.

Dynamic Mechanical Analysis
Shear moduli data were collected using a Texas Instruments Dynamic Mechanical Analysis 850 equipped with shear plates. Shear tests were performed on sample squares of length 7 mm at a frequency of 1 rads$^{-1}$ across a temperature range from 0 ˚C to 80 ˚C. For the aligned nematic samples, the orientation of the director relative to the oscillations is denoted by planar parallel (0˚), planar perpendicular (90˚), or homeotropic (90˚ out of plane).

Contact Angle

To collect the contact angle information, a Nikon AF-S NIKKOR lens (focal length 24mm) was fitted to a Nikon D3100 digital camera and the images evaluated. The LCEs were suspended in a bespoke mechanical strain rig, [20] and strained in steps of $\epsilon = 0.05$. A volume of 3 µL of deionized water was dropped onto the surface at each strain step and allowed to freely form a droplet. The contact angle is defined as the angle between the tangent of the droplet and the surface it is suspended upon. [18]

Acknowledgements
Conceptualization, AS & HFG; methodology, AS; resources, HFG; data curating, AS; formal analysis, AS; original draft preparation, AS; writing—review and editing, AS, JM & HFG; supervision, DM, JM & HFG; funding acquisition, HFG. All authors have read and agreed to the published version of the manuscript.

Received: [[will be filled in by the editorial staff]]
Revised: [[will be filled in by the editorial staff]]
Published online: [[will be filled in by the editorial staff]]

**Geometry-Dependent Adhesion in Transparent, Monodomain Liquid Crystal Elastomers**

ToC figure

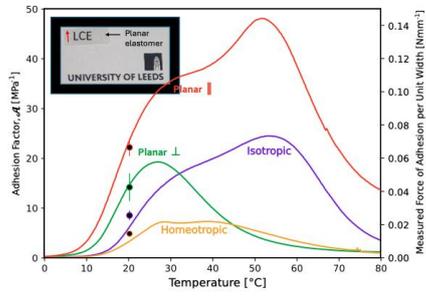